\documentclass[apj]{emulateapj}

\shorttitle{CR spectrum} \shortauthors{Katz et al.}

\usepackage{amsmath}
\usepackage{amssymb}

\newcommand\Mesz{M\'esz\'aros~}
\newcommand{\Nesc}{N_{\text{esc}}}
\newcommand{\xesc}{x_{\text{esc}}}
\newcommand{\ECR}{E_{\text{CR}}}
\newcommand{\vep}{\varepsilon}
\newcommand{\ep}{\epsilon}
\newcommand{\eV}{\mbox{eV}}
\newcommand{\sref}{\S~\ref}
\newcommand{\al}{\alpha}
\newcommand{\de}{\delta}
\newcommand{\bt}{\beta}
\newcommand{\erg}{\mbox{ erg}}
\newcommand{\cm}{\mbox{ cm}}
\newcommand{\haf}{\frac{1}{2}}

\begin{document}

\title{The spectrum of Cosmic Rays escaping from relativistic shocks}
\author{Boaz Katz\altaffilmark{1}, Peter \Mesz\altaffilmark{2} and Eli Waxman\altaffilmark{1}}%
\altaffiltext{1}{Weizmann Institute of Science, Rehovot, Israel} \altaffiltext{2}{Dpt. of Astronomy \& Astrophysics; Dpt. of Physics; Center for Particle
Astrophysics,Pennsylvania State University, University Park, PA 16802, USA}


\begin{abstract}
We derive expressions for the time integrated spectrum of Cosmic Rays (CRs) that are accelerated in a decelerating relativistic shock wave and escape ahead of the shock.
It is assumed that at any given time the CRs have a power law form, carry a constant fraction of the energy $E$ of the shocked plasma, and escape continuously at the
maximal energy attainable. The spectrum of escaping particles is highly sensitive to the instantaneous spectral index due to the fact that the minimal energy,
$\vep_{\min}\sim \Gamma ^2 m_pc^2$ where $\Gamma$ is the shock Lorentz factor, changes with time. In particular, the escaping spectrum may be considerably harder than the
canonical $N(\vep)\propto \vep^{-2}$ spectrum. For a shock expanding into a plasma of density $n$, a spectral break is expected at the maximal energy attainable at the
transition to non relativistic velocities, $\vep\sim 10^{19}(\ep_{B}/0.1)(n/1{\rm cm}^{-3})^{1/6}(E/10^{51}{\rm erg})^{1/3}\eV$ where $\ep_{B}$ is the fraction of the
energy flux carried by the magnetic field. If ultra-high energy CRs are generated in decelerating relativistic blast waves arising from the explosion of stellar mass
objects, their generation spectrum may therefore be different than the canonical $N(\vep)\propto\vep^{-2}$.
\end{abstract}



\section{Introduction}
\label{sec:intro}

Cosmic rays (CRs) are widely believed to be accelerated in collisionless shock waves in various systems \citep[see e.g.][]{Blandford87, Axford94}. In particular,
supernova remnant (SNR) shocks propagating subrelativistically are thought to accelerate protons up to $\sim 10^{15}$ eV \citep[see e.g.][]{Ginzburg69}, while non or
trans-relativistic internal shocks or ultra-relativistic external shocks of the jets of active galactic nuclei \citep[AGNs, see e.g.][]{Berezinsky08} or gamma-ray bursts
\citep[GRBs,][]{Waxman95, Vietri95, Milgrom95} may be the sources of Ultra High Energy CRs (UHECRs) up to the GZK energies of $\sim$ few~$\times10^{20}$ eV.

In cases where the shocked material expands considerably before releasing the CRs, most of the cosmic rays loose most of their energy by adiabatic losses before escaping.
Thus, the instantaneous spectrum of CRs at a given time and the integrated spectrum of the CRs escaping from the system may be different.

In this paper, we write down simple analytic expressions for the spectrum of escaping CRs from a relativistic decelerating shock wave assuming that at any given time (a)
CRs escape at the maximal energy attainable at that time, (b) the CRs have a power law energy spectrum, and (c) the CRs carry a constant fraction of the shocked thermal
plasma. We show that even under these simple, widely acceptable assumptions, the resulting spectrum may be non trivial and in particular significantly harder than the
canonical $N(\vep)\propto\vep^{-2}$ spectrum.

In section \sref{sec:Assumptions} we formulate our basic assumptions and give general expressions for the resulting flux. In section \sref{sec:Result} we focus on energy
conserving blast waves expanding in a uniform medium. The implications of the results are discussed in \sref{sec:Discussion}. Expressions for the escaping CR spectrum
expected in other scenarios, including power law density profiles, radiative shocks, and a jet injection with constant luminosity, are derived in \sref{sec:other}.

\section{Basic model and assumptions}\label{sec:Assumptions}
\label{sec:simpl}

Consider CRs that are accelerated by an expanding shock wave which is decelerating due to interaction with an external medium.

\subsection{Assumptions}
\noindent (i) The basic assumption that we make is that at any given time (or shock radius $R$), particles escape at the maximal energy $\vep_{\max}$ attainable at that
time. Specifically, it is assumed that $\vep\Nesc(\vep)$, the number of CRs ejected within a logarithmic energy interval around $\vep$, is similar to the number of
particles in the system, $\vep N(\vep,R)$, at the radius $R$ for which the maximal energy $\vep_{\max}$ is equal to $\vep$,
\begin{equation}\label{eq:BasicAssumption}
\vep\Nesc(\vep)\sim\vep N(\vep,R|_{\vep_{\max}=\vep}).
\end{equation}
Here and everywhere else in this letter, all quantities are measured in the observer frame. This assumption is expected to be true whenever the acceleration is limited by
the finite size of the system \citep[for a recent discussion see e.g.][]{Caprioli09}. Note that for a relativistic shock with Lorenz factor $\Gamma$, only particles that
move in the shock propagation direction to within an angular deviation of $1/\Gamma$, can outrun the shock and escape. This does not introduce a further correction
however, due to the fact that all particles in the shocked region, including the thermal and accelerated particles, are beamed in the observer frame to an angular
separation of $~1/\Gamma$ \citep[for examples of the expected angular distribution of CRs in relativistic shocks see e.g.][]{Kirk87,Bednarz98,Kirk00,Keshet05}.

For concreteness we further make the following assumptions.\\ \noindent (ii)  At any given radius $R$, the energy spectrum of CRs is a power law,
$N(\vep)\propto\vep^{-2-x}$ for $\vep_{\min}<\vep<\vep_{\max}$, with $x>0$ and with total energy $\ECR$. The spectrum can be expressed as
\begin{equation}
\vep^2N(\vep)=x\ECR\left(\frac{\vep}{\vep_{\min}}\right)^{-x},
\end{equation}
 where we assumed that $(\vep_{\max}/\vep_{\min})^{-x}\ll1$.\\

\noindent (iii) The minimal, maximal and total cosmic ray energies are power law functions of the radius,
\begin{equation}
\vep_{\min}\propto R^{-\al_{\min}},~~ \vep_{\max}\propto R^{-\al_{\max}},~~ \ECR\propto R^{-\al_{E}}.
\end{equation}

\subsection{Resulting spectra}
Under the above assumptions, the spectrum of escaped particles is given by
\begin{equation}\label{eq:xescDef}
\vep^2\Nesc(\vep)\propto\vep^{-\xesc}
\end{equation}
with
\begin{equation}\label{eq:GeneralPowerlaw}
\xesc=x-(\al_{\min}x+\al_E)/\al_{\max}. 
\end{equation}
For the case $\al_{min}=0$, this reduces to equation (28) of \citet{Ohira09}. 

Equation $\eqref{eq:GeneralPowerlaw}$ is valid for $x>0$. We note that for the limiting value, $x=0$, a logarithmic correction is introduced to the spectrum of escaped
CRs, due to the logarithmic dependence of the total CR energy on $\vep_{\min,\max}$, $\ECR=\log(\vep_{\max}/\vep_{\min})\vep^2N(\vep)$. For $x<0$, the energy carried by
the CRs is dominated by the particles with largest energies, $\vep=\vep_{max}$. The resulting escaped spectrum satisfies $\xesc=\al_E/\al_{max}$ and is not sensitive to
the form of the instantaneous spectrum.

\section{Energy conserving blast waves}\label{sec:Result}
We next apply the above general result to simple cases, focusing on a blast wave of fixed energy which starts off ultra-relativistic, and then decelerates to non
relativistic velocity. The non relativistic and relativistic stages are analyzed in \sref{sec:NR} and \sref{sec:UR} respectively. The spectrum resulting from the
combination of the two stages is addressed in \sref{sec:Discussion}.

\subsection{Non-relativistic expansion}\label{sec:NR}

Perhaps the simplest case to consider is the case of constant CR minimal energy $\vep_{\min}$ and total energy $\ECR$ ($\al_E,\al_{\min}=0$). This is expected in the
Sedov-Taylor phase of non relativistic blast waves. The energy in CRs is dominated by the relativistic CRs, $\vep>m_pc^2$, and is commonly assumed to be a constant
fraction $f$ of the total energy $E$ in the system. Hence $\vep_{\min}\sim m_pc^2$ and $\ECR=fE$, both independent of the shock radius. Eq. \eqref{eq:GeneralPowerlaw}
implies that the escaped spectrum is the same as the instantaneous spectrum,
\begin{equation}\label{eq:NonRel}
\xesc=x, \quad{\rm i.e.}\quad \Nesc(\vep)\propto\vep^{-2-x}.
\end{equation}
This result can be obtained directly from Eq. \eqref{eq:BasicAssumption} by noting that the entire spectrum, up to the maximal energy $\vep_{\max}(R)$, is independent of
radius.

\subsection{Ultra-relativistic expansion}\label{sec:UR}
We next consider the case of ultra-relativistic expansion. Below and in \sref{sec:other} we make the following assumptions in addition to assumptions (i)-(iii) above:\\
\noindent (iv) The minimal CR energy is the energy of the thermal particles,
\begin{equation}\label{eq:Emin}
\vep_{\min}\sim \vep_{\text{th}}\sim \Gamma^2 m_pc^2;
\end{equation}
\noindent (v) The CR pressure is a radius independent fraction $f_{\text{CR}}$ of the momentum flux in the shock frame, $p_{\text{CR}}\propto \Gamma^2\rho$, implying
\begin{equation}\label{eq:ECR}
\ECR\propto R^3\Gamma^2\rho;
\end{equation}
\noindent (vi) The maximal CR energy is the maximal energy of CRs that are confined by a fluid rest frame magnetic field $B_{\text{rest}}$ (equivalent to Diffusive Shock
Acceleration in the Bohm limit),
\begin{equation}\label{eq:Emax}
\vep_{\max}\sim eE_{\text{obs}}\de R\sim eB_{\text{rest}}R \propto e(\Gamma\rho^{1/2})R,
\end{equation}
where $E_{\text{obs}}\sim \Gamma B_{\text{rest}}$ is the electric field in the observer frame corresponding to a magnetic field $B_{\text{rest}}\sim (8\pi\Gamma^2\rho
c^2\ep_B)^{1/2}$ in the rest frame of the shocked plasma, assumed to carry a constant fraction $\ep_B$ of the momentum flux, and $\de R\sim R/\Gamma$ is the maximal
distance that a cosmic ray can propagate along the electric field in the observer frame.

\subsubsection{Ultra-relativistic impulsive expansion}

We next consider a Blandford- McKee \citep[energy conserving;][]{Blandford76} shock expanding into a uniform medium. By assumption $\al_E=0$. Using equations
\eqref{eq:Emin}-\eqref{eq:Emax}, $\al_{\max}=1/2$ and $\al_{\min}=3$ are obtained. Using \eqref{eq:GeneralPowerlaw} we find $\xesc=-5x$, or
\begin{equation}\label{eq:UltraRel}
\Nesc(\vep)\propto\vep^{5x-2}.
\end{equation}

\section{Discussion}\label{sec:Discussion}

The spectral index of escaped CRs, given by Eq.~\eqref{eq:UltraRel}, may be surprisingly hard if the instantaneous spectrum (equ.[2]) is softer than a flat spectrum, i.e.
$x>0$. The basic reason for this is the fact that the minimal CR energy $\vep_{\min}\sim \Gamma^2m_pc^2$ changes with radius much faster than the maximal energy, implying
that at later times (corresponding to lower escaped energies) there is much less CR energy at the escaping, high end of the spectrum.

For example, consider the commonly assumed Diffusive Shock Acceleration (DSA) mechanism \citep{Krymskii77, Axford77, Bell78, Blandford78}, which for ultra-relativistic
shocks with isotropic, small-angle scattering, leads to an instantaneous spectrum $N\propto \vep^{-20/9}$ \citep{Keshet05,Bednarz98,Kirk00}. Using Eq.
\eqref{eq:UltraRel}, this implies a very hard escaping spectrum, $\Nesc\propto \vep^{-8/9}$. We emphasize that DSA has not been shown to work based on first principles
and even if it does, the resulting spectrum is sensitive to the scattering mechanism \citep[e.g.][]{Keshet05,Keshet06} which is poorly understood. In fact, the
instantaneous spectrum does not necessarily need to have a power-law form \citep[e.g.][]{Caprioli09,Ohira09} or to have a constant spectral
index\citep[e.g.][]{Ellison04}.

Note that an instantaneous spectrum that is flat $N\propto \vep^{-2}$ \citep[e.g.][]{Katz07} or harder \citep[e.g.][and others]{Keshet06, Stecker07} can be obtained in
specific particle acceleration models. These models, however, are different from the possible hard spectrum of escaping particles discussed above. In fact, such hard
instantaneous spectra lead to a flat escaped spectrum $\Nesc(\vep)\propto \vep^{-2}$ [see discussion following eq. \eqref{eq:GeneralPowerlaw}]. The important point is
that the escaping spectrum is highly uncertain, very sensitive to the acceleration mechanism, and may be considerably harder than $\Nesc(\vep)\propto \vep^{-2}$.

Once the shock decelerates to non relativistic speeds, the integrated escaping spectrum changes its form. In the subsequent Sedov-Taylor phase, the minimal CR energy does
not change any more, $\vep_{\min}\sim m_pc^2$, and the escaping spectrum will be similar to the instantaneous spectrum [see Eq. \eqref{eq:NonRel}]. Note that in the
extreme case in which CRs carry most of the energy, and assuming they are accelerated by non-linear DSA \citep[for a recent review see][]{Malkov01}, the instantaneous
spectrum may not be a power law and correspondingly the escaping spectrum is different than that given here \citep[e.g.][]{Caprioli09,Ohira09}. In any case, a significant
break is expected in the escaping spectrum at an energy $\vep_\ast$ equal to the maximal energy achievable at the transition from relativistic to non-relativistic shock
velocities. Using Eq. \eqref{eq:Emax} for $\Gamma\bt\sim1$, the break is expected at
\begin{align}
\vep_{\ast}&\sim eB_{\text{rest}}R\sim (8\pi\rho c^2 \ep_B)^{1/2}[3E/(4\pi\rho c^2)]^{1/3}\cr &\sim 10^{19}\ep_{B,-1}n_0^{1/6}E_{51}^{1/3}\eV,
\end{align}
where we have taken, as an example, a gamma-ray burst source with  $E=10^{51}E_{51}\erg$, $\rho=n~m_p$ is the ambient density with  $n=n_0\cm^{-3}$, and
$\epsilon_B=0.1\epsilon_{B,-1}$. The spectral index above this energy will be $-2-\xesc=-2+5x$, as given by Eq. \eqref{eq:UltraRel}.  We note that the energy range where
$\xesc=-5x$ is limited between $\vep_\ast$ and $\max(\vep)\sim\vep_{\max}(\Gamma_0)$ where $\Gamma_0$ is the initial Lorentz factor. The maximal energy scales with
$\Gamma$ as $\vep_{\max}\propto\Gamma^{1/3}$ resulting in a narrow energy range $\vep_\ast\lesssim\vep\lesssim\Gamma_0^{1/3}\vep_\ast$. While this range depends on model
parameters, it does point to a possible mechanism for achieving a hardening of the spectrum at the high end of the CR spectrum.

We conclude that the spectrum of UHECRs, may be different than $\vep^{-2}$, and possibly considerably harder, e.g. if they originate from a relativistic decelerating
blast-wave resulting from the explosion of a stellar mass object.

\section*{Acknowledgments}
BK \& EW acknowledge support by ISF, AEC and Minerva grants. PM acknowledges support from NSF PHY-0757155 grant.
\newline

\newpage

\appendix
\section{Other particular cases}\label{sec:other}
Consider an ultra-relativistic shock, which expands into a medium with a density profile
\begin{equation}\label{eq:rho}
\rho\propto R^{-\alpha_\rho}
\end{equation}
with a bulk Lorentz factor
\begin{equation}
\Gamma\propto R^{-\al_{\Gamma}}.
\end{equation}

Using Eqs. \eqref{eq:Emin}-\eqref{eq:Emax} and \eqref{eq:rho} we find:
\begin{equation}
\al_{\min}=2\al_{\Gamma},~\al_{\max}=\al_{\Gamma}+\haf\al_{\rho}-1,~\text{and}~~\al_{E}=2\al_{\Gamma}+\al_{\rho}-3.
\end{equation}
Substituting this in equation \eqref{eq:GeneralPowerlaw} we find
\begin{equation}\label{eq:AxescGeneral}
\xesc=3-2\al_{\Gamma}-\al_{\rho}+
               \frac{(2-\al_{\rho})+2\al_{\Gamma}}{(2-\al_{\rho})-2\al_{\Gamma}}x.
\end{equation}
For a uniform density distribution, $\al_{\rho}=0$, and for a wind-like density distribution, $\al_\rho=2$, this reduces to
\begin{equation}\label{eq:AxescUniform}
\xesc=\frac{3-2\al_{\Gamma}}{\al_{\Gamma}-1}+\frac{1+\al_{\Gamma}}{1-\al_{\Gamma}}x,
\end{equation}
and
\begin{equation}\label{eq:AxesWind}
\xesc=\frac{1-2\al_{\Gamma}}{\al_{\Gamma}+1}-x
\end{equation}
respectively. The escaping spectrum is then
\begin{equation}
N_{esc}(\vep) \propto \vep^{-2-x_{esc}},
\end{equation}
which is flatter then $-2$ for $x_{esc}<0$. The expected spectral indices $\xesc$, for various physical scenarios, are provided in table \ref{tab:inds}.

\begin{table}
\caption{Spectral indexes of escaping CRs for different relativistic scenarios}\label{tab:inds}
\begin{center}
\begin{tabular}{|l|l|l|l|l|}
\hline Scenario & Conserved quantity &$\al_{\Gamma}$ & $\xesc ~~~[\rho\propto R^0]$ & $\xesc~~~~[\rho\propto R^{-2}]$\\ \hline Impulsive, Energy conserving & $E\sim
\Gamma^2\rho R^3$ & $(3-\al_{\rho})/2$& $-5x$ & $-x$\\ \hline Impulsive, Radiative& $M\Gamma\sim \Gamma\rho R^3$ & $3-\al_{\rho}$& $-3/2-2x$&$-1/2-x$\\ \hline Continuous
injection L$=$const. & $L\sim 4\pi\rho\Gamma^4c^3R^2$& $(2-\al_{\rho})/4$& -4+3x& 2-x\\ \hline
\end{tabular}
\end{center}
The value of the spectral index $\xesc$ is given by Eqs. \eqref{eq:AxescUniform} and \eqref{eq:AxesWind} for the various scenarios. 
$N(\vep)\propto \vep^{-2-\xesc}$.
\end{table}



\end{document}